\pdfoutput=1
\def\bfseries{\fontseries\bfdefault\selectfont\boldmath}
\def\itshape{\fontshape\itdefault\selectfont\let\mathrm=\mathit}
\documentclass[journal]{IEEEtran}
\usepackage{cite}
\usepackage{amssymb}
\usepackage{graphicx}
\usepackage{lineno}

\newcommand{\mm} {\ensuremath{\mathrm{mm}}}

\newcommand{\MeV}{\ensuremath{\mathrm{MeV}}}

\newcommand{\MHz}{\ensuremath{\mathrm{MHz}}}
\newcommand{\kHz}{\ensuremath{\mathrm{kHz}}}

\begin{document}
\linenumbersep2pt
\linenumberwidth3pt
\title{Proton and Neutron Irradiation Tests of Readout Electronics of the
  ATLAS Hadronic Endcap Calorimeter}
\author{Sven Menke on behalf of the ATLAS Liquid Argon Calorimeter Group
\thanks{Manuscript received November 16, 2012.}
\thanks{S.~Menke is with the Max-Planck-Institut f{\"u}r Physik, M{\"u}nchen, Germany (e-mail: menke@mpp.mpg.de).}%
}


\maketitle
\pagestyle{empty}
\thispagestyle{empty}

\begin{abstract}
  The readout electronics of the ATLAS Hadronic Endcap Calorimeter
  will have to withstand the about ten times larger radiation
  environment of the future high-luminosity LHC (HL-LHC) compared to
  their design values. The GaAs ASIC which comprises the heart of the
  readout electronics has been exposed to neutron and proton radiation
  with fluences up to ten times the total expected fluences for ten
  years of running of the HL-LHC. Neutron tests were performed at the
  NPI in {\v R}e{\v z}, Czech Republic, where a $36\,\MeV$ proton beam
  is directed on a thick heavy water target to produce neutrons.  The
  proton irradiation was done with $200\,{\rm MeV}$ protons at the
  PROSCAN area of the Proton Irradiation Facility at the PSI in
  Villigen, Switzerland.  In-situ measurements of S-parameters in both
  tests allow the evaluation of frequency dependent performance
  parameters -- like gain and input impedance -- as a function of the
  fluence. The linearity of the ASIC response has been measured
  directly in the neutron tests with a triangular input pulse of
  varying amplitude.  The performance measurements and expected
  performance degradations under HL-LHC conditions are presented.
\end{abstract}

\section{Introduction}
\IEEEPARstart{T}{he} electronics for the Hadronic Endcap Calorimeter
(HEC)~\cite{art:HEC} of the ATLAS Experiment~\cite{art:ATLAS} has been
designed to withstand ten years of operation of the LHC at
center-of-mass energies of $\sqrt{s}=14\,{\rm TeV}$ with peak
luminosities of ${\cal L} = 10^{34}\,{\rm cm}^{-2}\,{\rm s}^{-1}$
corresponding to an integrated luminosity of $\int{\cal L} =
1000\,{\rm fb}^{-1}$. The foreseen upgrade to the high-luminosity
version of the LHC (HL-LHC) exceeds these limits by up to an order of
magnitude. The expected radiation levels for integrated luminosities
of $3000\,{\rm fb}^{-1}$ in the region of the amplifier electronics
located inside the cryostat of the HEC are $5.1\times10^{13}\,{\rm
  h}\,{\rm cm}^{-2}$ for hadrons above $20\,{\rm MeV}$, including
safety factors of $5$ in total to account for simulation
uncertainties, while the corresponding number for $1\,\MeV$ equivalent
neutrons in silicon is $4.1\times 10^{14}{\rm n}/{\rm cm}^2$, about an
order of magnitude larger. The expected total ionization dose (again
evaluated for silicon and including the same safety factor of $5$) is
$6.2\,{\rm kGy}$. Proton and neutron irradiation tests have been
performed to evaluate the limits of the current readout electronics
under HL-LHC conditions.

In addition to the in-situ measurements during the irradiation which
took place at room temperature, the neutron irradiated boards were
re-measured $3$ months after the neutron tests, immersed in liquid
nitrogen to simulate the conditions inside the ATLAS endcap cryostat
where the readout electronics of the HEC are located.

\section{Proton Tests}\label{sec:proton}
\begin{figure}[!t]
\centering
\resizebox{0.49\textwidth}{!}{\includegraphics{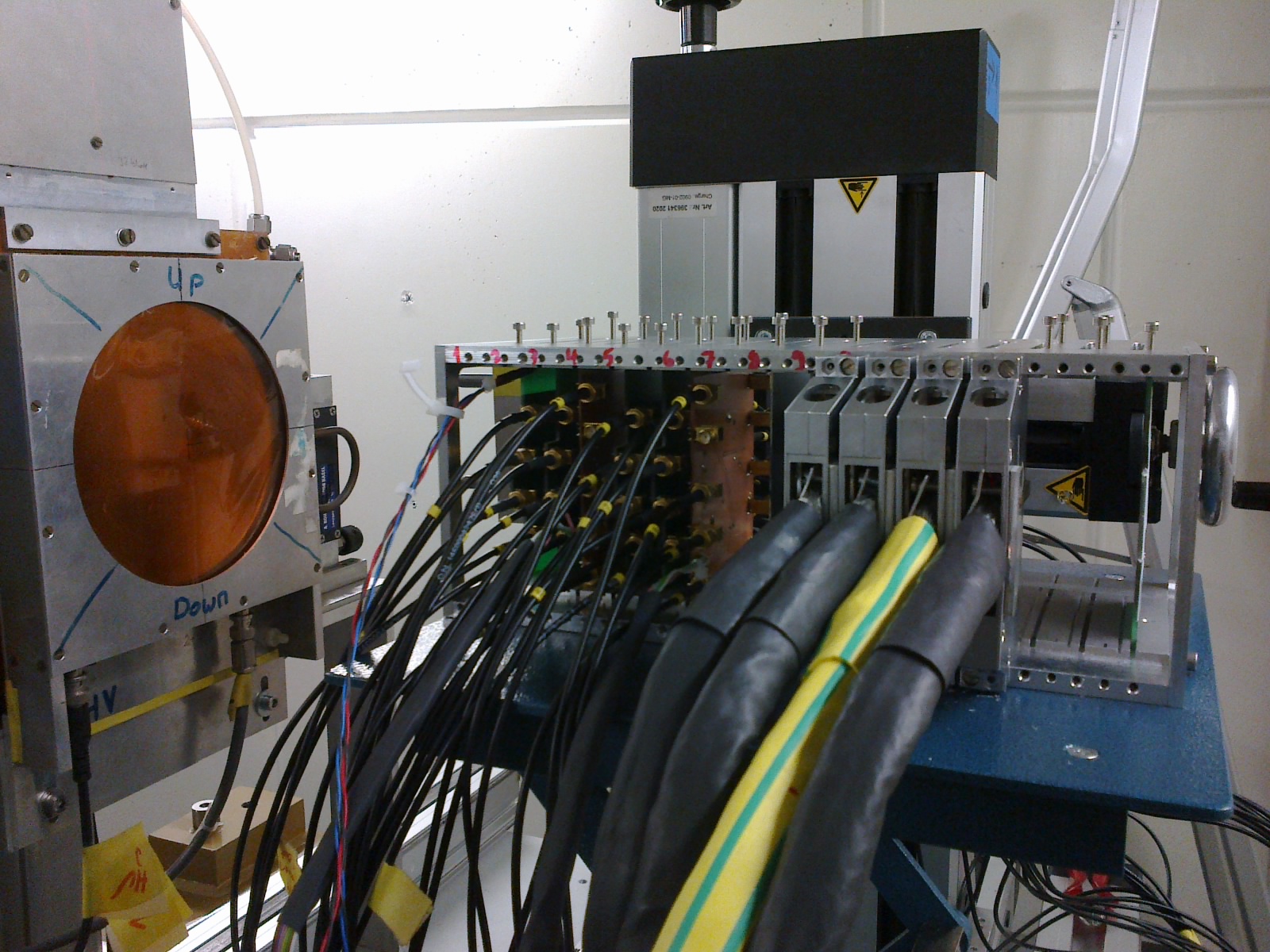}}
\caption{Picture of the test setup at PSI prior to final
  positioning. The proton beam comes from the left. An ionization
  chamber is visible as large aluminum square with a round hole on the
  left hand side. The test boards for the HEC readout electronics are
  mounted in the first 6 slots of the aluminum frame in the picture
  center.}
\label{fig:PIF_setup}
\end{figure}

A $198.9\,{\rm MeV}$ proton beam of $2.71\,{\rm nA}$ at the Proton
Irradiation Facility (PIF) at PSI with a narrow beam profile of
$\sigma_{x} \simeq 7\,{\rm mm}$ and $\sigma_{y} \simeq 9\,{\rm mm}$ at
the position of the devices under test has been used to evaluate the
radiation hardness against hadrons up to a fluence of
$2.6\times10^{14}\,{\rm p}\,{\rm cm}^{-2}$ after $22.05\,{\rm h}$ of beam time. 
A mount with 14 equidistant slots with a slot spacing of $1.7\,{\rm
  cm}$ has been placed in the beam line. The $6$ boards closest to the
beam window housed transistors in SiGe bipolar and Si CMOS FET
technology and the GaAs FET based ASICs (BB96)~\cite{art:BB96}. The
BB96 ASICs are currently used in ATLAS for the HEC readout. Slots $7$
and $14$ were equipped with radiation diodes for dose measurements,
and slots $8-13$ with test structures for the HEC power supplies.  A
picture of the setup prior to final alignment is shown in
Figure~\ref{fig:PIF_setup}.

The beam position and width has been measured with sheets of EBT2
radiation film placed for $5\,$minutes in the beam at $6$ different
positions.
The total beam flux was measured and calibrated by the PIF group with
ionization chambers. The raw ionization chamber current measurements
in $1\,{\rm s}$ intervals and the conversion factor to the proton beam
current was used to calculate the accumulated proton charge as
function of time.  Together with the beam width and position
measurements from the radiation films the fluence for the devices
under test in each slot could be derived.
The alignment of the setup was done prior to irradiation by means of
two lasers indicating the horizontal and vertical planes containing
the beam direction. The laser system was however misaligned for the
horizontal plane which led to an offset of $(6.5\pm0.5)\,{\rm mm}$ in
$y$, while the positioning of the boards themselves limited the
precision in $x$ to about $\pm 1\,{\rm mm}$.
Taking into account the small variations of the beam profile along the
beam line, the offset of the beam in $y$, and the expected losses due
to nuclear reactions in the devices under test, the flux was found to
be the same for all slots.

\begin{figure}[!t]
\centering
\resizebox{0.49\textwidth}{!}{\includegraphics{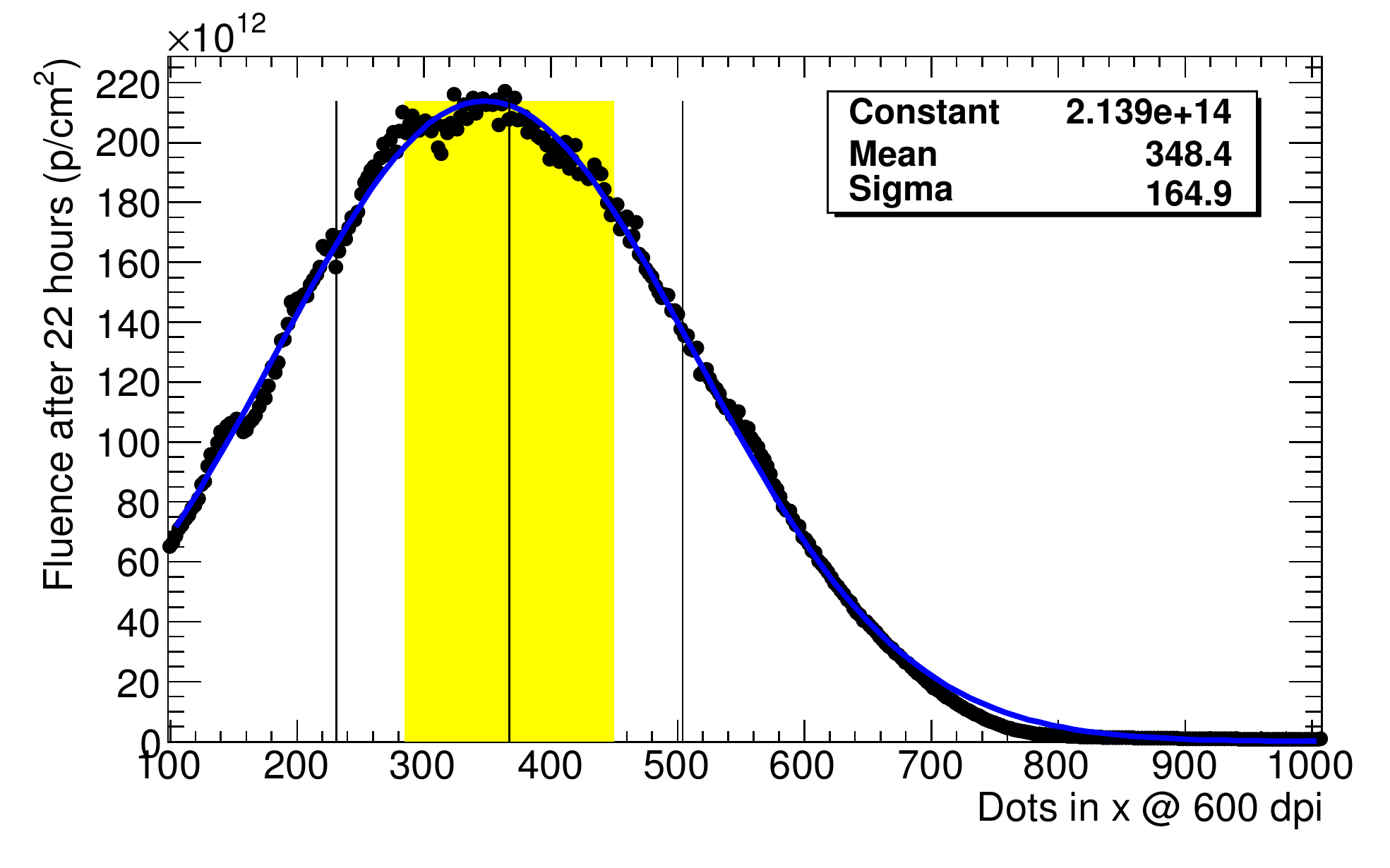}}
\resizebox{0.49\textwidth}{!}{\includegraphics{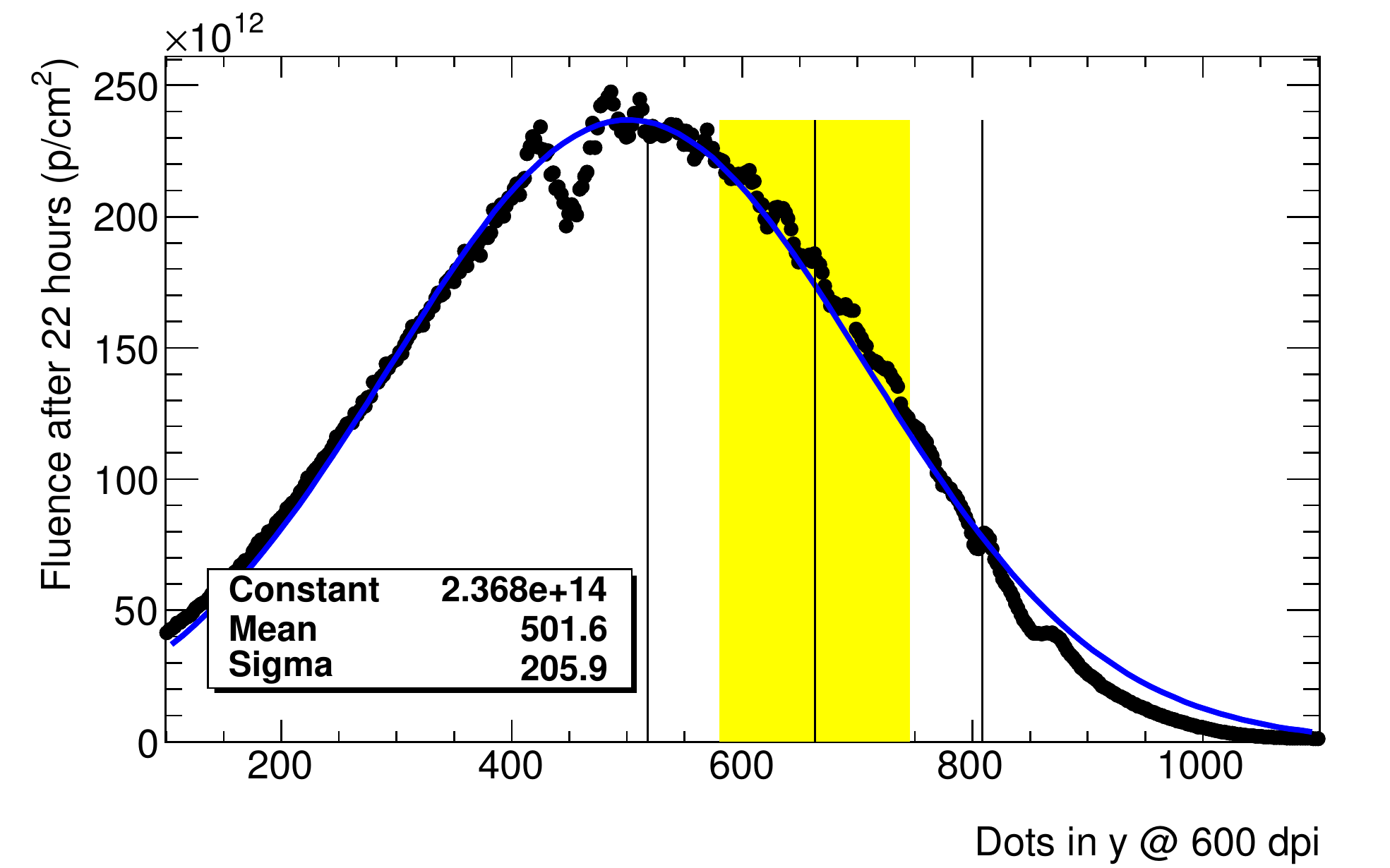}}
\caption{Horizontal (top) and vertical (bottom) profiles of the proton
  beam at PIF behind slot 3 normalized to the total fluence after
  $22\,{\rm hours}$ of beam time. The data points indicate the
  measurements from the radiation film. Superimposed are Gaussian
  fits. The yellow bands depict the position of the ASIC and the
  vertical lines the center and boundaries of the housing. The film
  was scanned with a resolution of $600\,$dots per inch (dpi). The
  Gaussian parameters for mean and width are also given in dpi.}
\label{fig:PIF_film_15}
\end{figure}

Figure~\ref{fig:PIF_film_15} shows the horizontal and vertical
profiles of the proton beam as measured by one of the radiation
films. Dips in the data distribution are caused by the soldered legs
of housings in the upstream slots. The superimposed Gaussian fits show
that the profiles are roughly Gaussian in the area of the ASIC.

\section{Neutron Tests}\label{sec:neutron}
\begin{figure}[!t]
\centering
\resizebox{0.49\textwidth}{!}{\includegraphics{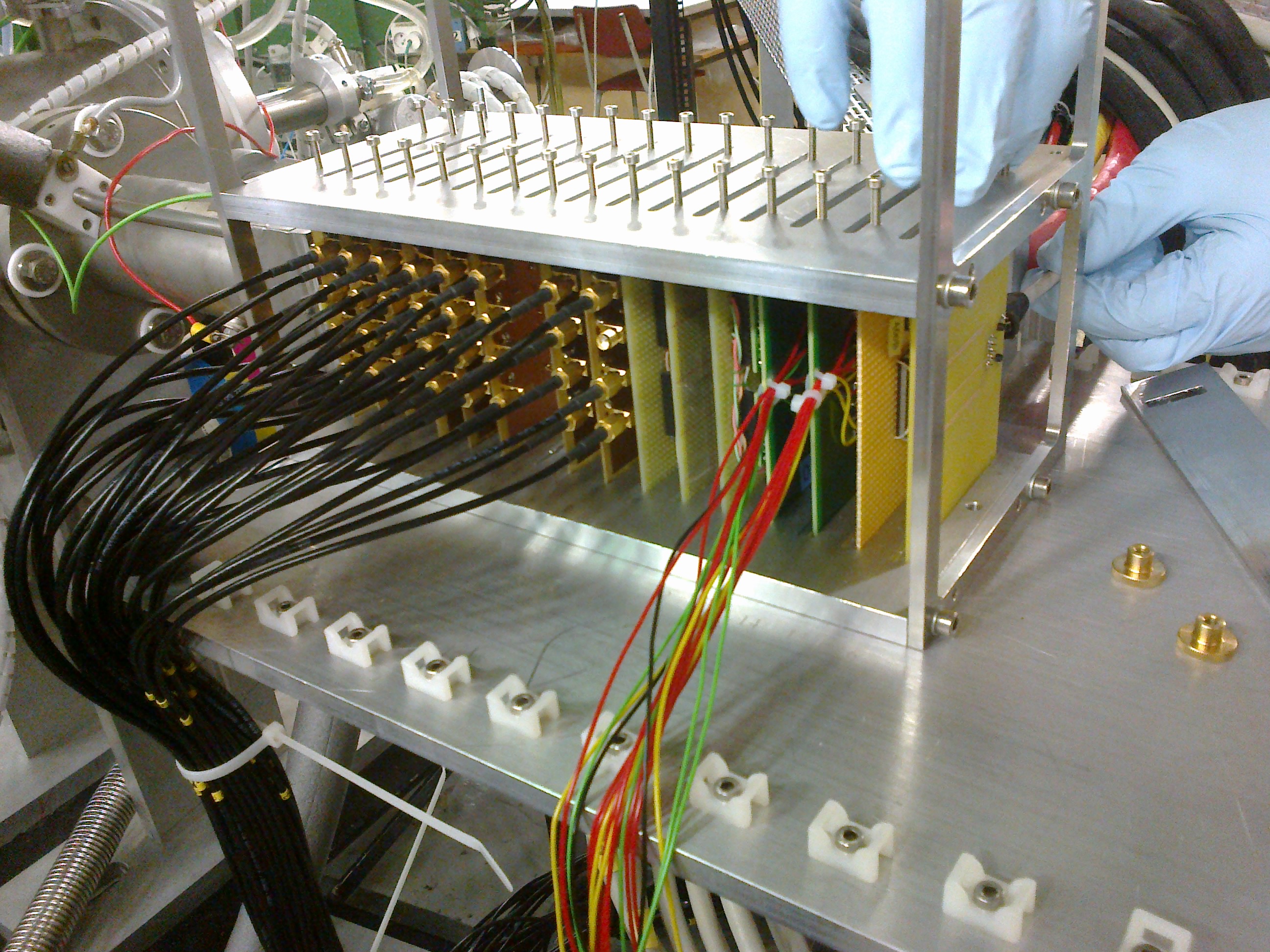}}
\caption{Picture of the test setup at NPI. The neutrons leave the
  ${\rm D}_2{\rm O}$ target at the the upper left part of the picture
  and travel to the right.  The test boards for the HEC readout
  electronics occupy the first 6 slots and slots 8 and 9 of the
  aluminum frame in the picture center.}
\label{fig:Rez_setup}
\end{figure}

Neutron tests have been performed with a $12-14\,\mu{\rm A}$ and
$36\,\MeV$ proton beam of the variable-energy cyclotron U-120 at NPI
in {\v R}e{\v z} hitting a $16\,\mm$ thick ${\rm D}_2{\rm O}$ target
to produce a neutron spectrum with $E_{\rm kin} < 32\,\MeV$ and
$\langle E_{\rm kin}\rangle \simeq 14\,\MeV$~\cite{art:Bem}.
Similar to the setup in the proton test, $16$ equidistant slots with
distances of $1.7\,{\rm cm}$ between adjacent board centers have been
equipped and placed in front of the $3\,\mm$ steel plate enclosing the
target. The alignment was done prior to mounting the target and the
test boards by burning small holes with the proton beam in two
calibration boards placed instead of the test structures in the first
and last slot.
Slots 1--6, 8, and 9 were equipped with the BB96 GaAs ASICs, slots 7
and 11 with radiation diodes and activation foils and the remaining
slots with test structures for the HEC power supplies. The setup is
shown in Figure~\ref{fig:Rez_setup}.

The total collected proton charge in the test was $2.0\,{\rm C}$.
The neutron flux profile has been measured with sheets of EBT2
radiation film placed for $5\,$minutes at a reduced machine current of
$1.2\,\mu{\rm A}$ in front of slot 1 and behind all other slots with
the exception of slots 7,10, and 11.  During a dedicated fluence test
run in 2010 with radiation monitors in each slot the following
empirical relation of the $1\,\MeV$ equivalent neutron fluence in
silicon and the machine current as function of the slot number has
been found~\cite{art:Dannheim}:
\begin{equation}
  \textrm{fluence} = \frac{1.61\times10^{10}\,{\rm n}}{{\rm cm}^2\mu{\rm C}} \left(\textrm{slot} + \frac{\textrm{offset}\pm 2\,{\rm mm}}{17\,{\rm mm}}\right)^{-2.11}, 
\end{equation}\label{eq:rez_fluence}
where slot is the slot-number (1-16), offset is the distance of the
first slot from the target ($3\,{\rm mm}$) and the positions are known
with $\pm2\,{\rm mm}$ precision.

\begin{figure}[!t]
\centering
\resizebox{0.49\textwidth}{!}{\includegraphics{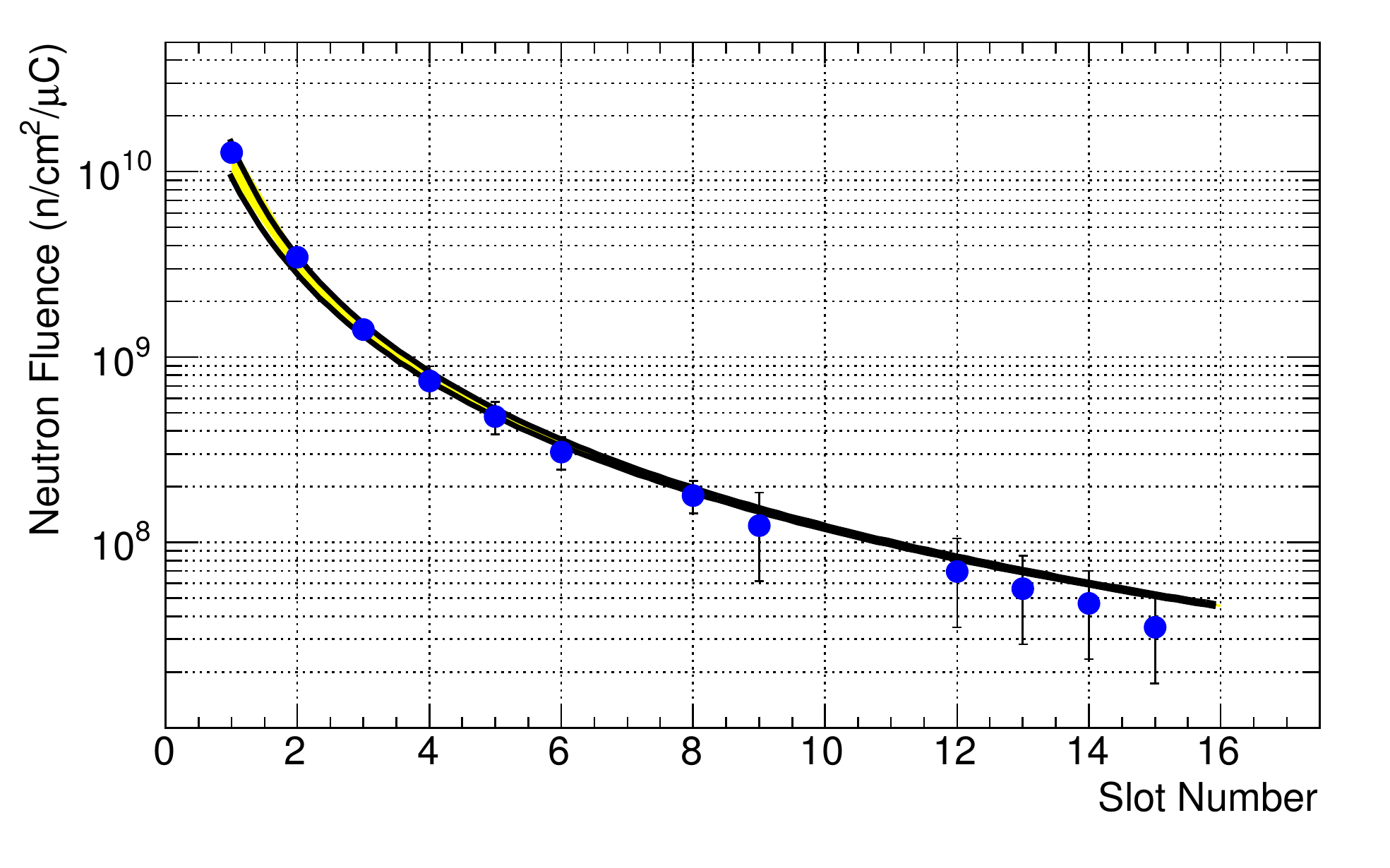}}
\caption{$1\,{\rm MeV}$ equivalent Neutron fluence in silicon as
  function of slot number. The data points depict relative dose
  measurements from radiation films normalized to the expected fluence
  in slot 3. The yellow band corresponds to the fluence curve measured
  2010 in a dedicated run with radiation monitors in each slot.}
\label{fig:rez_fluence}
\end{figure}

Figure~\ref{fig:rez_fluence} shows this fluence curve superimposed by
the relative fluence measurements from the radiation films, normalized
to the expected fluence in slot 3.

\section{S-Parameter Measurements}\label{sec:s-par}
During the irradiation the scattering parameters (S-parameters) of the
BB96 ASICs have been continuously measured to study their performance
as a function of the fluence up the maximum value at the end of the
irradiation. The S-parameters have been measured for the frequency
range $300\,\kHz$ to $100\,\MHz$ with a vector network analyzer (VNA)
connected via $40\,{\rm m}$ long cables to the devices under
test. Each cable pair has been individually de-embedded with the VNA
calibration utility moving the plane of calibration to the input
connectors of the boards. The most important parameters are the input
port reflection coefficient $S_{11}$, which can be translated into the
input impedance $Z_{\rm in} = 50\,\Omega \, (1+S_{11})/(1-S_{11})$,
for our case of vanishing feedback coefficient $S_{12}$ and low load
reflection, and the forward transmission coefficient $S_{21}$. The
product of the two gives the transimpedance gain in the frequency
domain. This can be interpreted as the input current to output voltage
signal amplification.
It was found that the effect of protons in the GaAs ASICs is similar
to that of neutrons with $\sim 5$ times larger fluence. Since the
expected ratio of neutron to proton fluence for the HL-LHC is $\sim
10$, the neutron tests provide the most limiting numbers.

\section{Pulse Measurements}\label{sec:pulse}
In the neutron test triangular voltage pulses similar to the
ionization currents in the LAr gaps in ATLAS with $10$ different
amplitudes between $0.7\,{\rm mV}$ and $70\,{\rm mV}$ and pulse
lengths of $500\,{\rm ns}$ have been applied over the $40\,{\rm m}$
cables at the BB96 ASICs and both the input and the output pulse have
been measured with an oscilloscope (the output pulse again after
$40\,{\rm m}$ cable).  These measurements were done alternatingly with the
S-parameter measurements and provide a more direct way of studying the
performance of the electronics. Good agreement with the S-parameter
measurements was found for the gain as a function of neutron fluence.
The largest expected ionization currents in ATLAS are $250\,\mu{\rm
  A}$ for a single pre-amplifier and about $500\,\mu{\rm A}$ for a
full chain of several pre-amplifiers and one summing amplifier. Up to
these values the linearity of the pre-amplifiers and the full system
was measured with the pulses. To convert from input voltage pulses to
currents the input impedance and cable damping is taken into account.
Thus the voltage range of $0.7\,{\rm mV}$ to $70\,{\rm mV}$
corresponds to an input current range of $10\,\mu{\rm A}$ to
$1000\,\mu{\rm A}$ for non-irradiated chips.

\section{Warm In-situ Results}\label{sec:warm} 
\begin{figure}[!t]
\centering
\resizebox{0.49\textwidth}{!}{\includegraphics{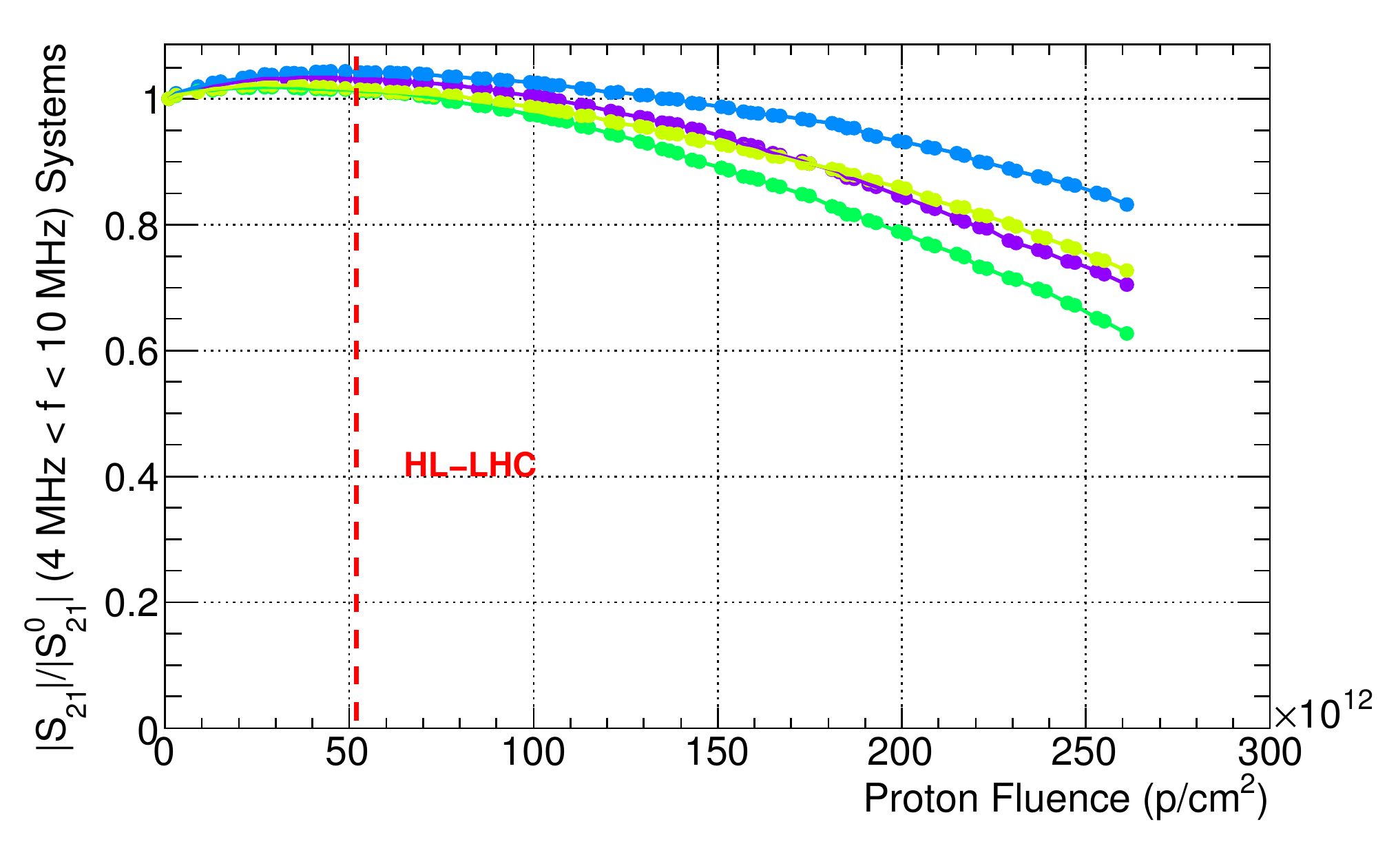}}
\resizebox{0.49\textwidth}{!}{\includegraphics{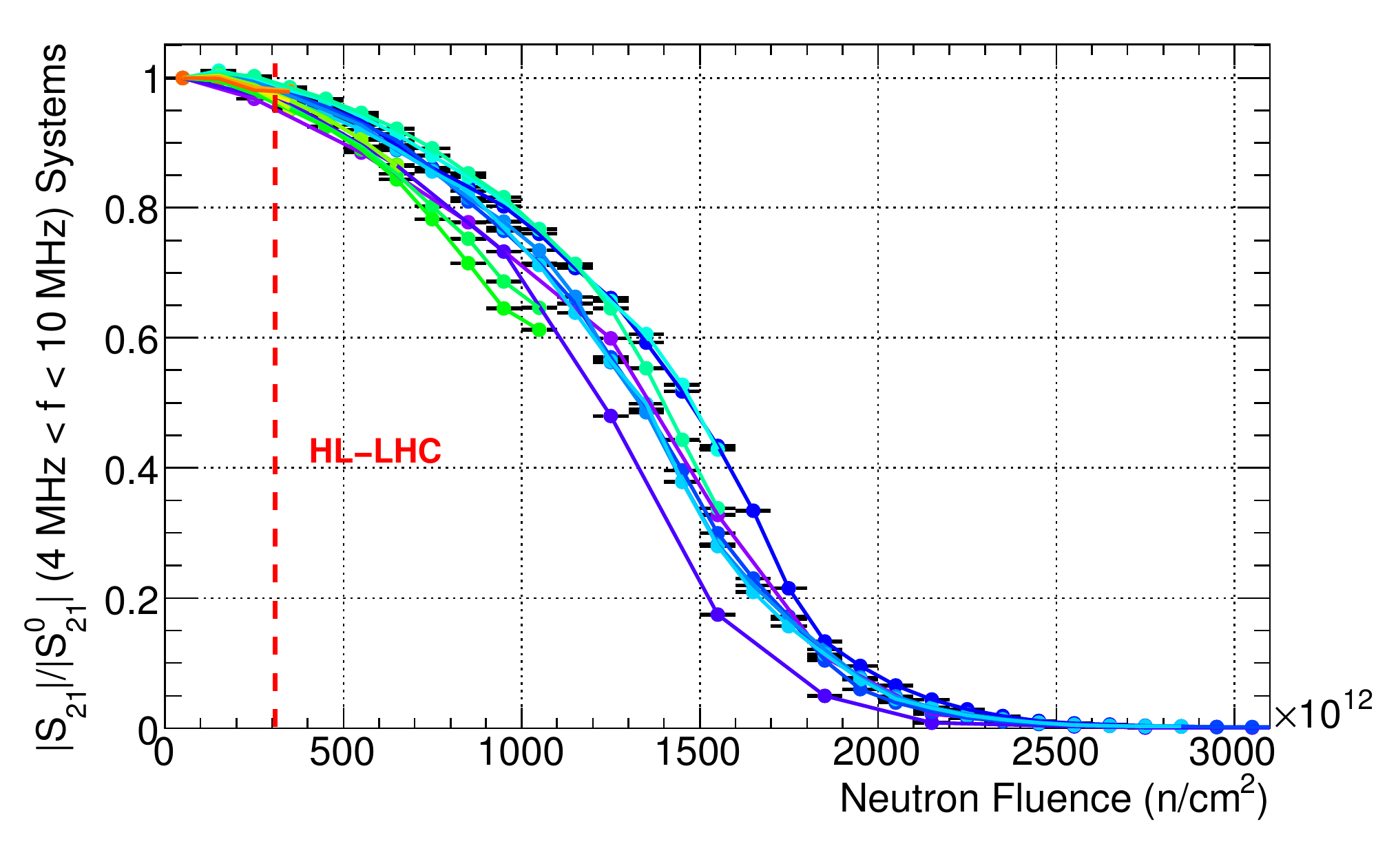}}
\caption{Forward transmission coefficient normalized to the value
  before irradiation and evaluated in the frequency range of the
  shaper electronics in ATLAS ($4\,{\rm MHz} < f < 10\,{\rm MHz}$) as
  measured in situ for a system of pre-amplifier and summing amplifier
  during the proton (top) and neutron (bottom) irradiations vs.~the
  fluence. The neutron fluence is given in $1\,{\rm MeV}$ equivalent
  for silicon. The red vertical lines indicate HL-LHC limits including
  a safety factor of $5$. In the neutron case the fluence is again the
  expected silicon $1\,{\rm MeV}$ fluence but corrected by the
  expected differences in ${\rm GaAs}$. The colored data points are
  from the various devices under test and the spread of the points
  indicate sample variations. Since for the neutron test the fluence
  varied with slot number the point density and maximal fluence per
  slot varies.}
\label{fig:S21_warm_SYS}
\end{figure}

The forward transmission coefficient normalized to the value before
irradiation and evaluated in the frequency range of the shaper
electronics in ATLAS ($4\,{\rm MHz} < f < 10\,{\rm MHz}$) as measured
in situ for the full systems consisting of one pre-amplifier and a
summing amplifier during the proton and neutron irradiations is shown
in Figure~\ref{fig:S21_warm_SYS} as function of the proton and
$1\,{\rm MeV}$ equivalent neutron fluence in silicon.
Figure~\ref{fig:S21_warm_SYS} shows that the device-to-device
fluctuations increase with irradiation and that the effect of protons
is about $4-5$ times larger on the voltage gain compared to that of
neutrons at the same fluence values. Input impedances were found to
stay close to $50\,\Omega$ and real throughout the entire irradiation
range for protons and up to $\simeq 4\times10^{14}\,{\rm n}/{\rm
  cm}^2$ for neutrons. The output impedance shows similar behavior and
stays roughly twice as long close to $50\,\Omega$. Beyond these limits
the impedances stay real but rise quickly resulting in non-matched
conditions in ATLAS, where $0.5-2\,{\rm m}$ $50\,\Omega$ cables connect
the detector and the pre-amplifiers and $\simeq10\,{\rm m}$
$50\,\Omega$ cables carry the signals from the ASIC output to the 
shaper electronics.
 
\section{Cold  Results}\label{sec:cold} 
About three months after the neutron irradiation tests at NPI in {\v
  R}e{\v z} the irradiated boards have been shipped to the lab in Munich
and re-tested with the same set of measurements as in situ in both
warm and cold (liquid nitrogen) conditions. The three front-most
boards received neutron fluences in excess of $2.8\times10^{15}\,{\rm
  n}/{\rm cm}^2$ (Si $1\,{\rm MeV}$ eq.~in {\v R}e{\v z}). From
Figure~\ref{fig:S21_warm_SYS} it can be seen that beyond
$2.1\times10^{15}\,{\rm n}/{\rm cm}^2$ the ASICs approach $0$ voltage
gain and therefore the three first boards are dead. The board in the
fourth slot was exposed to $1.6\times10^{15}\,{\rm n}/{\rm cm}^2$ and
this one plus the following four boards from slots 5,6,8, and 9, with
respective fluences of $10.0$, $6.9$, $3.8$, and
$3.0\times10^{14}\,{\rm n}/{\rm cm}^2$ are still operational in warm
conditions and still show the same performance as was measured at the
end of the run in situ at {\v R}e{\v z}. In addition to these 5
fluence points non-irradiated ASICs have been measured in warm and
cold as well to provide data at zero fluence.
\begin{figure}[!t]
\centering
\resizebox{0.49\textwidth}{!}{\includegraphics{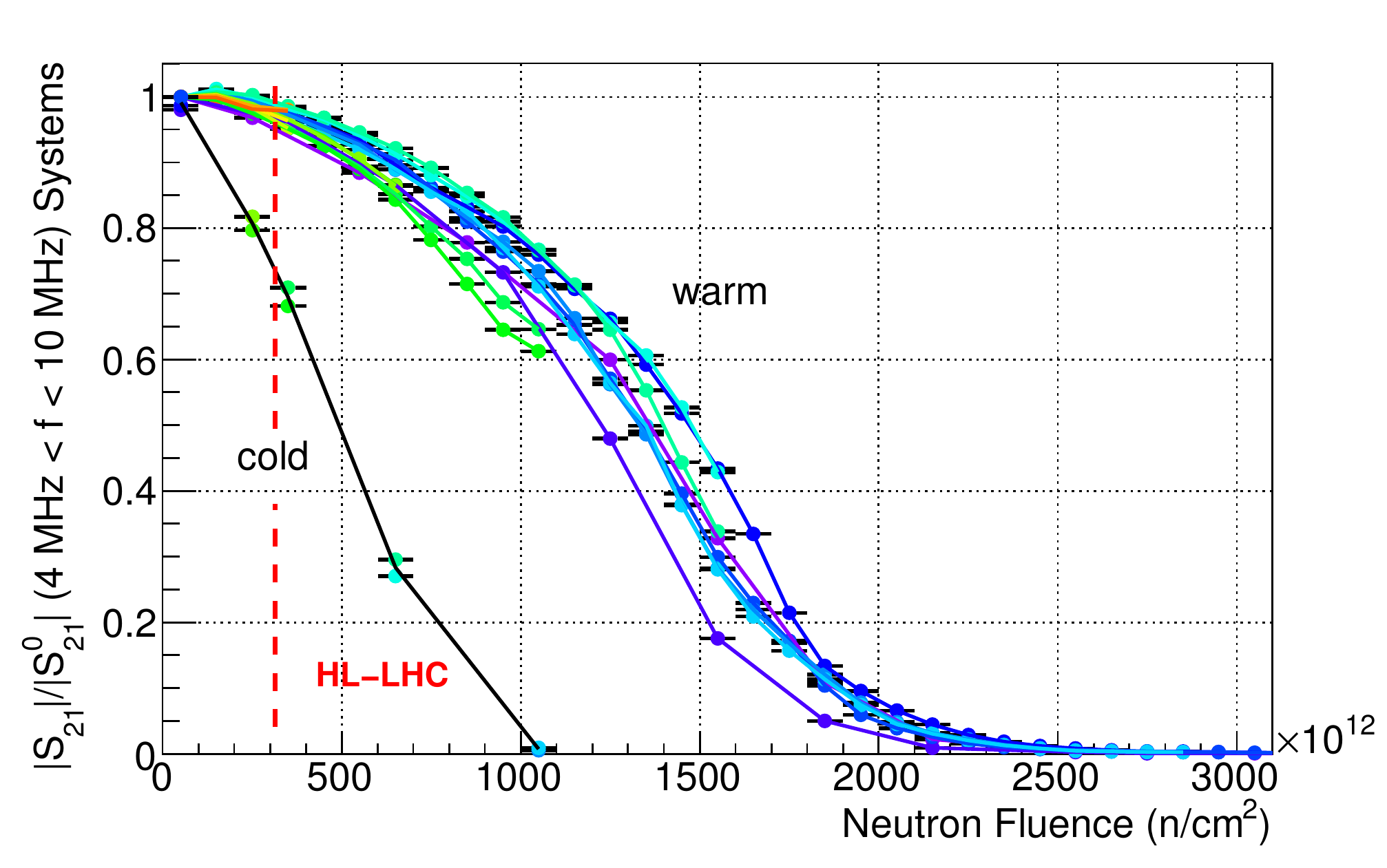}}
\resizebox{0.49\textwidth}{!}{\includegraphics{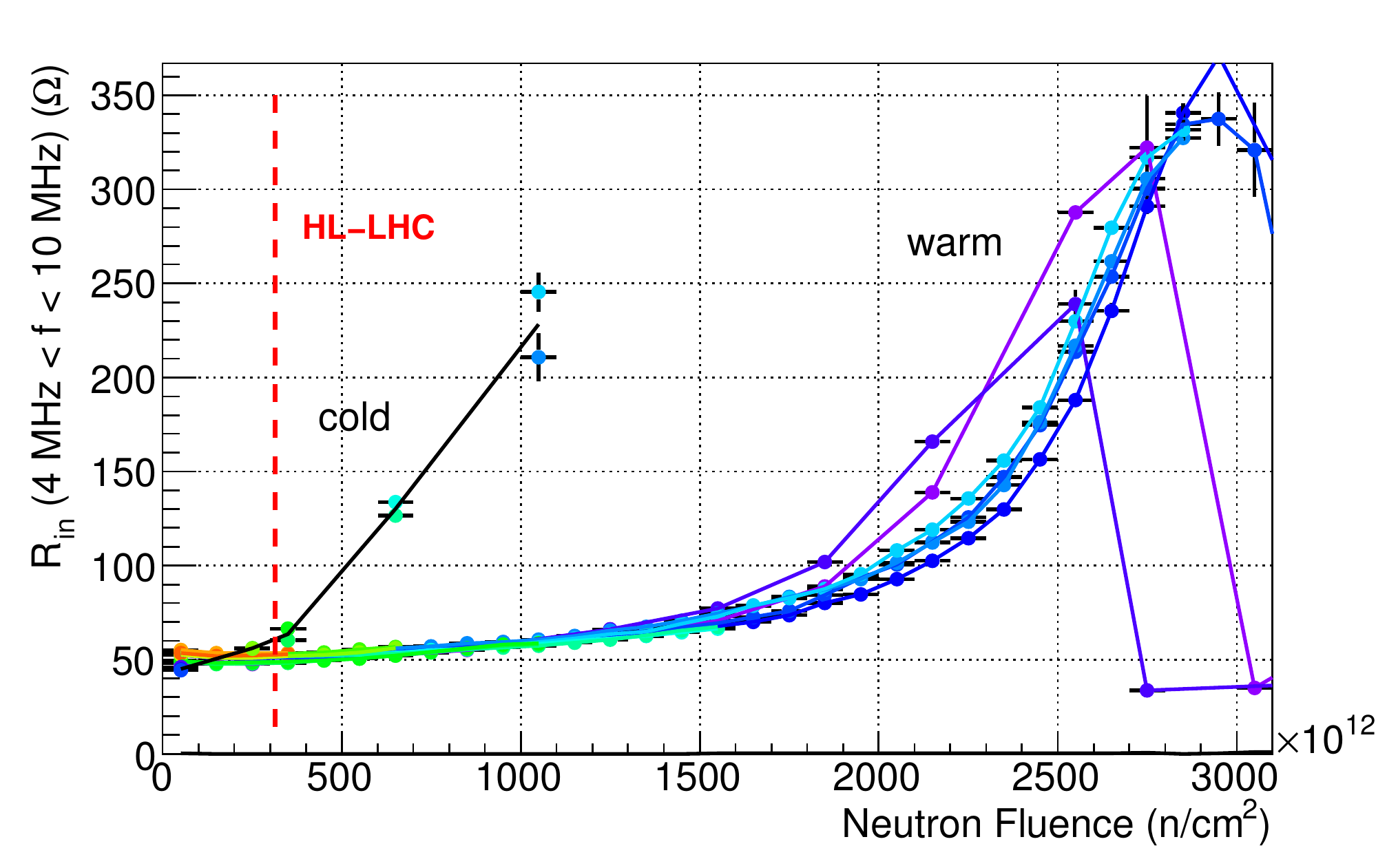}}
\caption{Forward transmission coefficient normalized to the value
  before irradiation (top) and input impedance (bottom) evaluated in
  the frequency range of the shaper electronics in ATLAS ($4\,{\rm
    MHz} < f < 10\,{\rm MHz}$) as measured in situ for systems of a
  pre-amplifier and a summing amplifier during the neutron
  irradiations in warm and three months after the irradiation in cold
  vs.~the fluence. The neutron fluence is given in $1\,{\rm MeV}$
  equivalent for silicon. The red vertical lines indicate the HL-LHC
  limit including a safety factor of $5$.}
\label{fig:S21_Rin_warm_cold_SYS}
\end{figure}
Immersed in liquid nitrogen to emulate the conditions inside the ATLAS
cryostat the performance is dramatically worse than in warm
conditions. Figure~\ref{fig:S21_Rin_warm_cold_SYS} shows the
normalized forward transmission coefficient and the input impedance
for the systems comprised of one pre-amplifier and a summing amplifier
as function of neutron fluence for both warm and cold conditions.  The
sparse data points on the left indicate the cold measurements, while
the denser groups of points extending to larger fluences are the
measurements in the warm. The performance in the cold is roughly
equivalent to that in warm at $3$ times larger fluences. The
performance degrades quickly beyond $3-4\times10^{14}\,{\rm n}/{\rm
  cm}^{2}$.

\begin{figure}[!t]
\centering
\resizebox{0.49\textwidth}{!}{\includegraphics{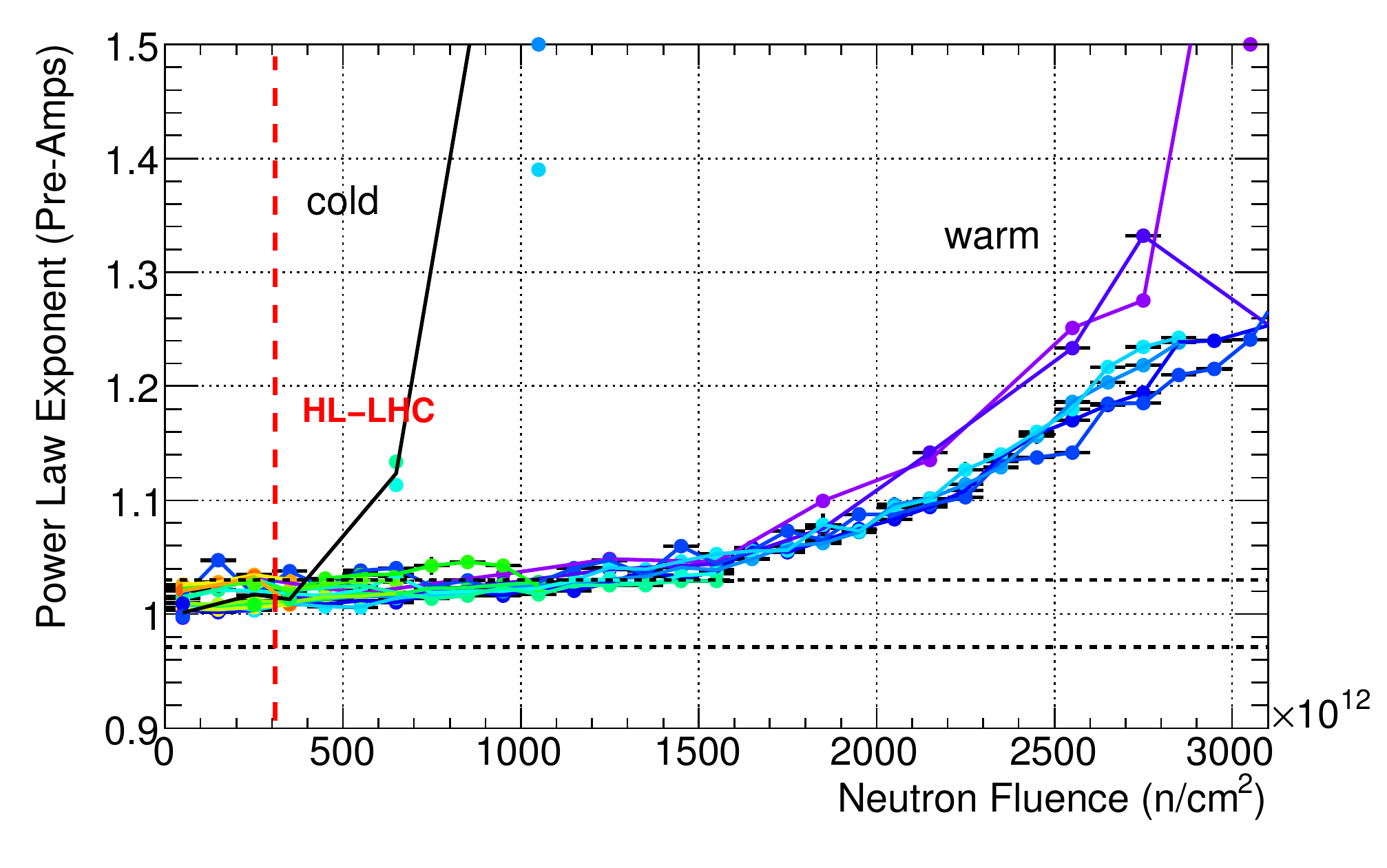}}
\caption{Measured non-linearity of the pre-amplifiers as function of
  neutron fluence in warm (dense data points extending to large
  fluences) and in cold (sparse data points at lower fluences).  The
  non-linearity is expressed in terms of an power-law
  exponent. Deviations from $1.0$ indicate non-linear
  behavior. Horizontal lines indicate exponent limits for a deviation
  of the output amplitude from linear expectation normalized to the
  maximum amplitude of $1\%$. The neutron fluence is given in $1\,{\rm
    MeV}$ equivalent for silicon. The red vertical line indicates the
  HL-LHC limit including a safety factor of $5$.}
\label{fig:pa_non_lin}
\end{figure}
The pulse measurements as described in section~\ref{sec:pulse} have
also been repeated with the already irradiated boards in warm and in
cold. The pulse amplitudes have been adjusted to lower levels to still
correspond to $10-1000\,\mu{\rm A}$ input currents for non-irradiated
ASICs for the short ($\sim 4\,{\rm m}$) cables used at MPP. The
increase of non-linearity of the response was found to be described by
a power-law relation of input ($\textrm{in}$) and output
($\textrm{out}$) amplitude of the form:
\begin{equation}
\textrm{out} = c\times\textrm{in}^x,
\end{equation}\label{eq:power_law}
where $c$ is a fluence dependent constant and $x$ the power-law
exponent. Values for $x$ deviating from $1.0$ indicate non-linear
behavior.

Figure~\ref{fig:pa_non_lin} shows this exponent as function of neutron
fluence in warm and cold. Like in the case of the forward transmission
coefficient the non-linearities for irradiated ASICs are worse in the
cold and correspond roughly to those in the warm at $3$ times larger
fluences. Since up to $16$ pre-amplifiers are summed by one summing
amplifier in ATLAS the non-linearities for the pre-amplifiers can not
be corrected.  An exponent of $1.03$ or $0.97$ would correspond to a
maximum deviation of the output amplitude from linear expectation
normalized to the maximum amplitude of $1\%$. This original
requirement limit for the non-linearity of the pre-amplifiers is
indicated by the horizontal lines in Figure~\ref{fig:pa_non_lin}.
Beyond neutron fluences of $3-4\times10^{14}\,{\rm n}/{\rm cm}^2$ the
non-linearity quickly rises above this level.

\begin{figure}[!t]
\centering
\resizebox{0.49\textwidth}{!}{\includegraphics{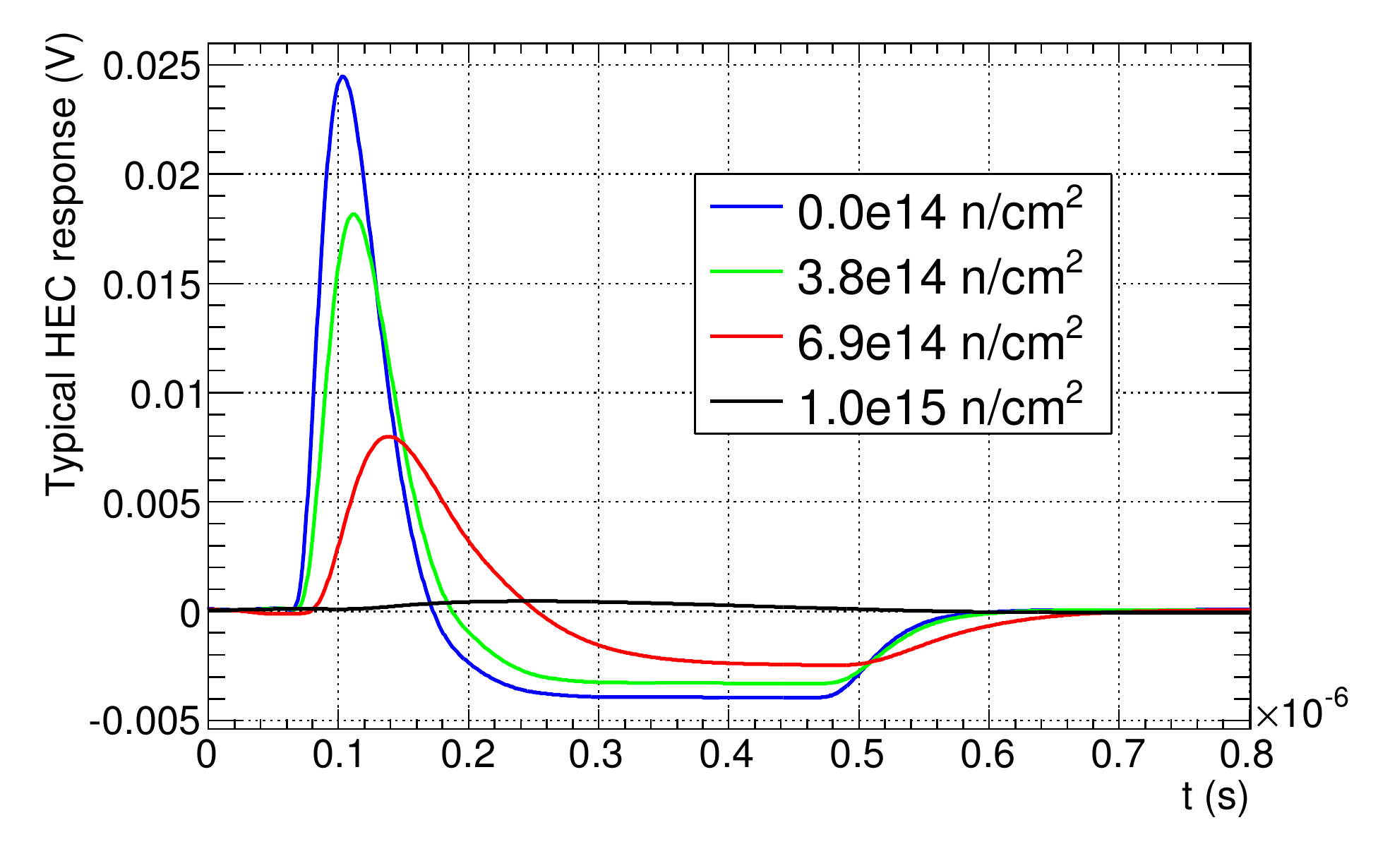}}
\caption{Simulated typical response to a $400\,{\rm ns}$ triangular
  ionization current in the HEC for one channel for 4 different
  neutron fluences. The fluence is given as $1\,{\rm MeV}$ equivalent
  fluence in silicon for the neutron spectrum at NPI in {\v R}e{\v
    z}.}
\label{fig:pulse}
\end{figure}

The degraded S-parameter measurements in the full frequency range of
$300\,{\rm kHz}-100\,{\rm MHz}$ have been used to predict the response
to a typical ionization signal in the HEC after the full electronics
chain for the HEC in
ATLAS~\cite{art:HEC-note-109}. Figure~\ref{fig:pulse} shows the
simulated response to a $400\,{\rm ns}$ triangular current in a
typical HEC channel at 4 different fluence values. It can be seen that
on top of the gain degradation the pulse shape worsens with increasing
fluence leading to a slower rise- and fall-time, a broader and more
delayed positive peak. 

\section{Fluence Conversion for GaAs}\label{sec:fluence} 
The neutron fluence values given in all the figures presented so far are
$1\,{\rm MeV}$ equivalent fluences evaluated for the neutron spectrum
at NPI in {\v R}e{\v z}~\cite{art:Bem}. In order to compare these
values to expectations for the HL-LHC the fluences have to be
converted for the $\textrm{GaAs}$ devices studied here first to the
corresponding $1\,{\rm MeV}$ equivalent fluence in $\textrm{GaAs}$
using the NPI neutron spectrum and the Kerma factor for neutrons in
silicon and gallium-arsenide~\cite{art:Kerma}. In a second step using
the expected neutron, proton and pion spectra in
ATLAS~\cite{art:ATLASRad} the $1\,{\rm MeV}$ equivalent neutron
fluence in $\textrm{GaAs}$ has to be converted back to $\textrm{Si}$
again using Kerma factors for the relevant radiation in both
$\textrm{Si}$ and $\textrm{GaAs}$. It was found that the conversion
from Si to GaAs at NPI energies is $1.82$ and from GaAs to Si at the
location of the ASICs in ATLAS $1/1.35$.  All measured neutron
fluences at NPI reported here can thus be converted to ATLAS
expectations by multiplying with $1.82/1.35$.

The proton-irradiated boards were not retrievable from PSI and thus 
cold measurements of the proton irradiated ASICs are not available.
Assuming that
\begin{enumerate}
\item the results in the cold would be worse compared to the warm
  in-situ measurements by the same factor of $3$ as observed in the
  neutron tests
\item the effect of $1\,{\rm p}/{\rm cm}^2$ with kinetic energy of
  $200\,{\rm MeV}$ in $\textrm{GaAs}$ according to the Kerma factor
  and specific energy loss~\cite{art:PDG12} can be split in $1.4\,{\rm
    n}/{\rm cm}^2$ $1\,{\rm MeV}$ equivalent neutron fluence in
  $\textrm{GaAs}$ and $465\,{\rm pGy}$ total ionization dose (TID)
\end{enumerate}
it is possible to convert the $\textrm{GaAs}$ numbers back to
$\textrm{Si}$ again and estimate the impact of the expected
$\textrm{Si}$ TID of $6.2\,{\rm kGy}$ in ATLAS at the HL-LHC, which
would be equivalent to $\sim10\%$ of the expected degradation due to
$1\,{\rm MeV}$ equivalent neutrons.
  
\section{Conclusions}\label{sec:conclusions} 
Irradiation tests of the current readout electronics of the Hadronic
Endcap Calorimeter of ATLAS with neutrons and protons with fluence
levels exceeding those relevant for the high luminosity upgrade of the
LHC have been performed. The most stringent limits have been observed
under neutron irradiation and the measurements at liquid nitrogen
temperatures show a $\sim3$ times faster performance degradation
compared to room temperature measurements.
 
The performance limit in the cold was reached at
$3-4\times10^{14}\,{\rm n}/{\rm cm}^{2}$ $1\,{\rm MeV}$ equivalent in
$\textrm{Si}$ for neutrons in {\v R}e{\v z}, where the forward
transmission coefficient reduces to $\simeq70\%$, the input
(output) impedance increases by $\simeq20\%$ ($\simeq40\%$) above the
nominal $50\,\Omega$, and the power-law exponent reaches
$\simeq1.03$. These levels correspond to $4-5.4\times10^{14}\,{\rm
  n}/{\rm cm}^{2}$ $1\,{\rm MeV}$ equivalent in $\textrm{Si}$ for
ATLAS at the position of the ASICs.

With the current safety factor of $5$ for the radiation level
simulations for ATLAS this limit will be reached at the HL-LHC. The
impact on the ATLAS physics performance of the degraded electronics
and upgrade options of the readout electronics for the HL-LHC are
currently under study.


\begin{thebibliography}{10}
\bibitem{art:HEC} D.~M.~Gingrich {\it et al.}, \emph{Construction,
  assembly and testing of the ATLAS hadronic end-cap calorimeter},
  \newblock JINST {\bf 2} P05005 (2007).

\bibitem{art:ATLAS} The ATLAS Collaboration, \emph{The
  ATLAS Experiment at the CERN Large Hadron Collider}, \newblock JINST
  {\bf 3} S08003 (2008).

\bibitem{art:BB96} J. Ban {\it et al.}, \emph{Cold Electronics for the Liquid
 Argon Hadronic End-cap Calorimeter of ATLAS}, \newblock
 Nucl.~Instr.\&\ Meth. {\bf A556}, 158-168 (2006).

\bibitem{art:Bem} P.~B{\'e}m {\it et al.}, \emph{The NPI cyclotron-based fast neutron facility}, 
\newblock International Conference on Nuclear Data for Science and Technology 2007, 555-558, http://dx.doi.org/10.1051/ndata:07598 (2007).

\bibitem{art:Dannheim} D.~Dannheim, \emph{n-flux measurement
{\v R}e{\v z} irradiation April 2010}, http://indico.mppmu.mpg.de/indico/getFile.py/access?contribId=3\&resId =1\&materialId=slides\&confId=834 (2010).

\bibitem{art:HEC-note-109} L.~Kurchaninow, \emph{Modeling of the HEC Electronics Chain}, ATLAS HEC-Note-109 (2001). 

\bibitem{art:Kerma} 
A.~Vasilescu and G.~Lindstroem, \emph{Displacement damage in silicon}, on-line compilation, http://sesam.desy.de/members/gunnar/Si-dfuncs.html,
\newblock
P.J.~Griffin {\it et al.}, SAND92-0094 (1993),
\newblock
A.~Konobeyev, J.~Nucl.~Mater. {\bf 186},117 (1992),
\newblock
G.P.~Summers {\it et al.}, IEEE NS {\bf 40}, 1372 (1993),
\newblock
M.~Huhtinen and P.A.~Aarnio, Nucl.~Instr.\&\ Meth. {\bf A335}, 580 (1993),
\newblock
A.M.~Ougouag {\it et al.}, IEEE NS {\bf 31}, 2219 (1990),
\newblock
A.L.~Barry {\it et al.}, IEEE NS {\bf 42}, 2104 (1995).

\bibitem{art:ATLASRad} S.~Baranov {\it et al.}, \emph{Estimation of Radiation Background, Impact on Detectors, Activation and Shielding Optimization in ATLAS}, \newblock ATL-GEN-2005-001 (2005).

\bibitem{art:PDG12} J.~Beringer {\it et al.} (Particle Data Group), \emph{2012 Review of Particle Physics}, \newblock Phys.~Rev. {\bf D86}, 010001 (2012).

\end{thebibliography}
\end{document}